\documentclass{v15cosmo}

\def\be{\begin{equation}}
\def\ee{\end{equation}}
\def\bea{\begin{eqnarray}}
\def\eea{\end{eqnarray}}

\newcommand{\Photo}{}

\begin{document}
\vspace*{4cm}
\title{QUARK-GLUON PLASMA: FROM ACCELERATOR EXPERIMENTS TO EARLY UNIVERSE}

\author{ P. ROSNET }

\address{Universit\'e Clermont Auvergne, Universit\'e Blaise Pascal, CNRS/IN2P3, \\ Laboratoire de Physique Corpusculaire, Clermont-Ferrand, France}

\maketitle\abstracts{
In the Big Bang scenario, the early Universe is characterized by the {\it particle era}, i.e. a Universe made of particles. 
This period connects both scales of fundamental physics: infinitesimally small and infinitely large.
So, particle physics and in particular experimental programs at accelerators can bring valuable inputs for the understanding of the early Universe and its evolution.
These proceedings discuss the impact of the Quantum ChromoDynamics phase transition experienced by the {\it particle era} in the expanding Universe, which is connected to the study of the Quark-Gluon Plasma produced in heavy-ion physics experiments.}

\section{Introduction}

In the last years, high-tech experiments have confirmed the two theoretical pillars of fundamental physics.
At the infinitesimaly scale, the Standard Model (SM) of particle physics was conforted by the discovery of a scalar boson, by the CERN~\footnote{CERN is the European Organization for Nuclear Research.} Large Hadron Collider (LHC) experiments~\cite{LHC-BEH}. 
This boson has all properties of the expected Brought-Englert-Higgs (BEH) boson. 
At the opposite infinitely large scale, the Standard Model of cosmology, or Lambda Cold Dark Matter ($\Lambda$CDM) model, was tested with increasing precision by the ESA~\footnote{ESA is the European Space Agency.} Planck mission~\cite{Planck} by measuring the temperature and polarization maps of the Cosmic Microwave Background (CMB).
Despite these successes, fundamental open questions have still no answers~\cite{NRC2003,Ellis2005} (here after is not an exhautive list).
Is there an unified theoretical framework to describe all known fundamental interactions: electroweak, strong and gravitational forces?
Why is the Universe geometry nearly Euclidean?
What is the nature of the dark energy (about 68\% of the content of the present Universe)?
What is the nature of the dark matter (about 27\% of the Universe content)?
What is the origin of the observed matter-antimatter asymmetry in the Universe?
What is the origin of the fluctuations observed in the CMB which can explain the formation of cosmological structures?

\begin{figure}
\begin{minipage}{0.5\linewidth}
\centerline{\includegraphics[width=0.9\linewidth]{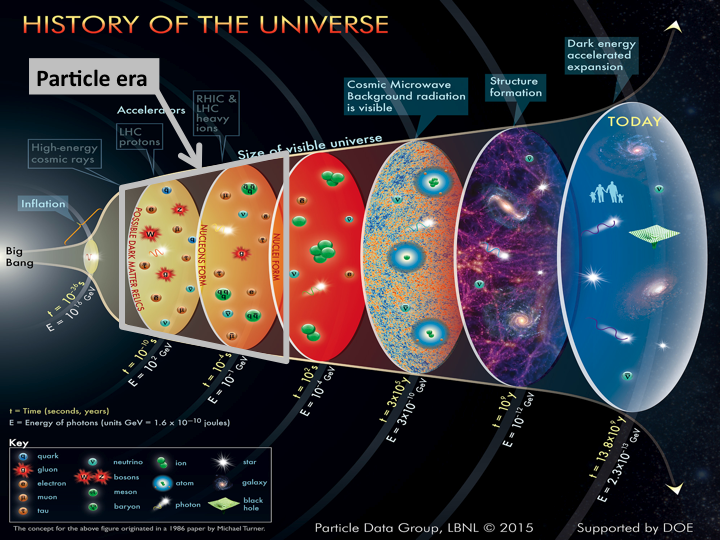}}
\end{minipage}
\hfill
\begin{minipage}{0.5\linewidth}
\centerline{\includegraphics[width=0.9\linewidth]{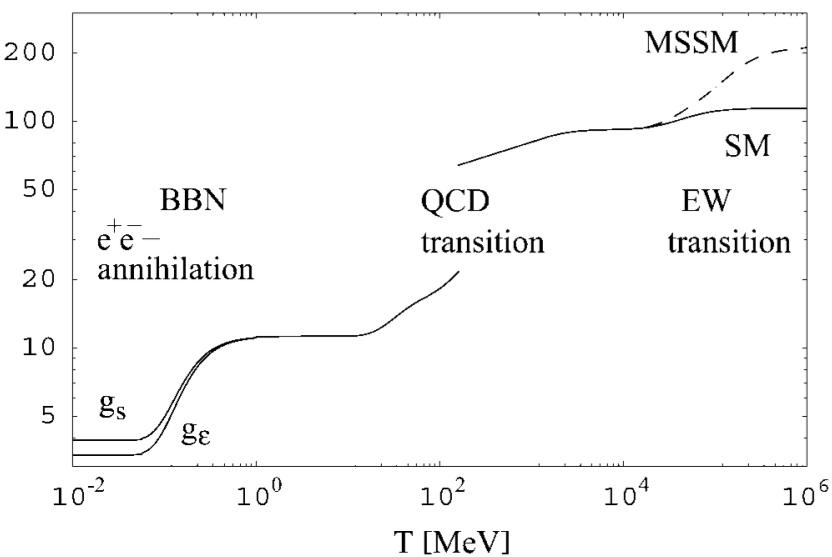}}
\end{minipage}
\caption[]{Left: scheme of the Universe history in the Big Bang scenario of the Standard Model of cosmology~\cite{PDG2015}.
The {\it particle era} is highlighted.
Right: evolution of the number of degrees of freedom (d.o.f.) as a function of the temperature: the full line is the prediction of the Standard Model of particle physics and the dashed line shows a minimal supersymmetric extension (after $e^+e^-$ annihilation, separation between energy density and entropy density d.o.f., denoted $g_\epsilon$ and $g_s$, respectively, is shown)~\cite{Schwarz2003}.}
\label{fig-Universe_SM}
\end{figure}

Even if the two theoretical frameworks are differents, the connexions between both scales are multiples and can be crucials because the {\it particle era} constitutes a critical period in the Big Bang scenario, as illustrated by Fig.~\ref{fig-Universe_SM} (left): between the conjectured inflationary period of the Universe and the formation of its large structures.
In this picture, fundamental interactions probed by particle accelerator experiments can answer some of the questions and bring valuable inputs for understanding the Universe evolution~\cite{Ellis2005}.

Of particular interest are the phase transitions experienced by the expanding Universe during the {\it particle era}.
Indeed, the number of degrees of freedom (d.o.f.) of particles evolves with the Universe temperature as illustrated in Fig.~\ref{fig-Universe_SM} (right): in a first approximation, the d.o.f. of a particle of mass $m$ can be excited only if $k T \ge m c^2$.
From Fig.~\ref{fig-Universe_SM}, it appears that the Universe underwent three main phase transitions during this period: 
\begin{itemize}
\item
the electroweak (EW) transition due to its spontaneous symmetry breaking where electroweak boson d.o.f. ($W^\pm$, $Z$ and $H$) could no longer be excited, 
\item
the Quantum ChromoDynamics (QCD) transtion due to hadron formation where quark and gluon d.o.f. disappear in favour of hadronic d.o.f., 
\item
the Big-Bang Nucleosynthesis (BBN) and the $e^+e^-$ annihilation transitions. 
\end{itemize}
Emphasis is given to the QCD phase transition which appears to be the strongest in term of change in the number of d.o.f.: $\Delta g_{\rm QCD} \approx 45$ (with respect to $\Delta g_{\rm EW} = 12$) from the total number of d.o.f. of the Standard Model of particle physics $g_{\rm SM} \approx 107$.

\section{The QCD phase transition: theoretical and experimental points of view}\label{sec-QCD}

\subsection{Asymptotic freedom of QCD}\label{subsec-QCDfreedom}

The Quantum ChromoDynamics (QCD) is the theoretical framework which describes the interaction between quarks and gluons.
There are 6 quark flavours (and the antiquarks) in the SM of particle physics with 3 color states each and 8 color gluon states.
The main property of QCD is the running of its coupling constant $\alpha_{\rm S}$ (S for strong force).  
This running shows that the strong coupling decreases when the energy increases.
At low energy, as the one characterizing confinment of quarks inside the hadrons, the strong coupling is close to unity ($\alpha_{\rm S} \sim 1$) and a perturbative development technique can not be applied.
In this case, the formal framework is the one proposed by lattice QCD (lQCD) consisting of a discretization of the space-time on a lattice which allows to perfom numerical calculations. 
At high energy, the strong coupling becomes far smaller than unity ($\alpha_{\rm S} \ll 1$) and perturbative QCD (pQCD) applied, leading to the property of asymptotic freedom in the limit of very high energies.

\subsection{The QCD phase diagram}\label{subsec-QCDphase}

\begin{figure}
\begin{minipage}{0.5\linewidth}
\centerline{\includegraphics[width=0.9\linewidth]{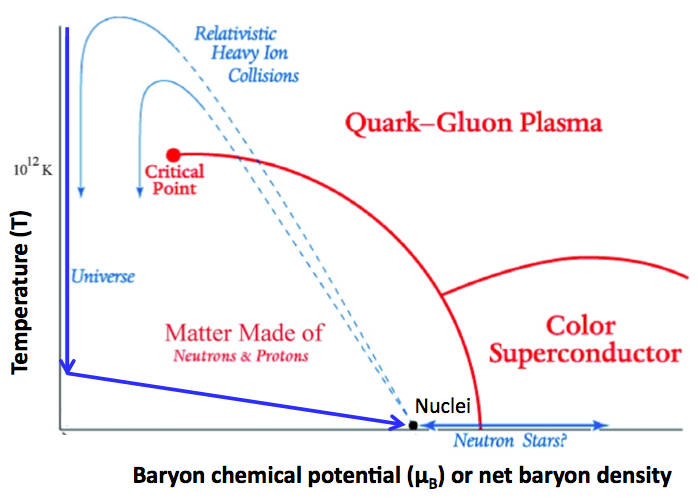}}
\end{minipage}
\hfill
\begin{minipage}{0.5\linewidth}
\centerline{\includegraphics[width=0.9\linewidth]{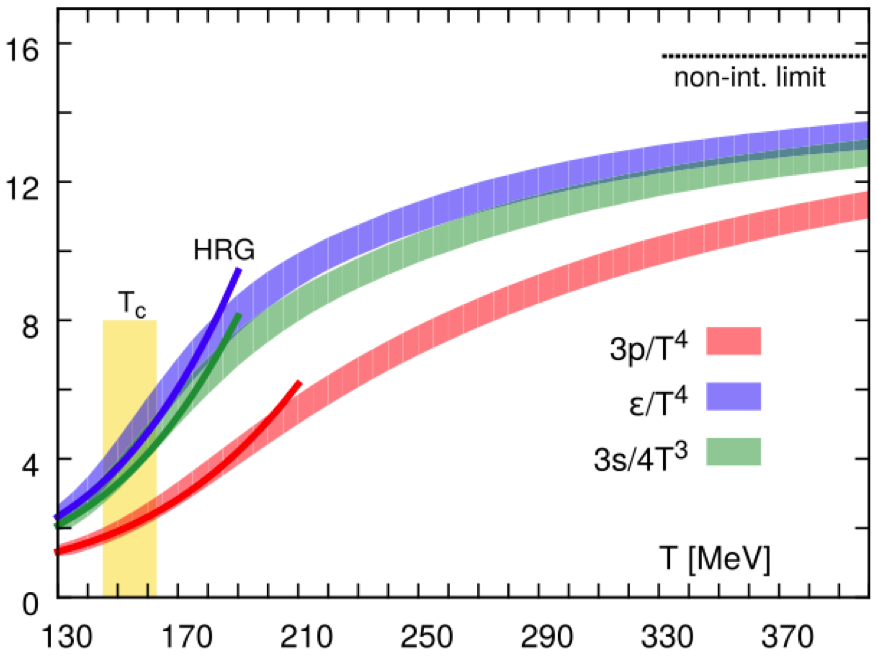}}
\end{minipage}
\caption[]{Left: QCD phase diagram in the plane temperature $T$ as a function of the baryon chemical potential $\mu_{\rm B}$ or the net baryon density (image adapted from Ref.~\cite{NRC2003} with Universe path from Ref.~\cite{MinkowskiKbana2002}).
Full red line represents first order phase transitions with a critical point.
Blue lines show expected condition in neutron star, but also evolution in the early Universe and in heavy-ion collisions.
Right: evolution of energy density $\epsilon$, pressure $P$ and entropy density $s$ as a function of temperature $T$ as predicted by lQCD at $\mu_{\rm B}=0$ (color bands) and comparison to the Hadron Relativistic Gas (HRG)~\cite{HotQCD2014}.
Upper right dotted line shows also the non-interacting gas limit of the deconfined state.}
\label{fig-QCD}
\end{figure}

The asymptotic freedom behaviour of QCD implies that at high energy density (i.e. high temperature and/or high net baryon density), quarks should be deconfined in a new state of matter~\cite{CabibboParisi1975}, called Quark-Gluon Plasma (QGP)~\cite{Shuryak1978}, but also a superdense color state expected in the core of neutron stars~\cite{CollinsPerry1975}.
The modern picture of the QCD phase diagram as predicted by lQCD is presented in Fig.~\ref{fig-QCD} (left) with a first order phase starting from high baryon chemical potential $\mu_{\rm B}$ ({\it i.e.} high net baryon density) and ending in a critical point. 
At $\mu_{\rm B}=0$, lQCD calculations shown in Fig.~\ref{fig-QCD} (right) indicates that the behaviour of energy density $\epsilon$ and pressure $P$ as a function of temperature $T$ is such that: (i) the phase transition is of cross-over type, (ii) $\frac{3P}{T^4} \neq \frac{\epsilon}{T^4}$ leading to the conclusion that the QCD Equation of State (EoS) does not correspond to the radiation EoS.
The recent extrapolation of lQCD calculation to the continuum limit at $\mu_{\rm B}=0$ gives for the temperature transition~\cite{HotQCD2012}: $T_{\rm QCD} = (154 \pm 9)$~MeV $\approx 1.8 \times 10^{12}$~K (temperature usually named critical temperature $T_{\rm C}$ in heavy-ion physics community), corresponding to a critical energy density~\cite{HotQCD2014} $\epsilon_{\rm QCD} = (0.18 - 0.50)$~GeV/fm$^3$.

The QCD phase transition is accompagnied by the spontaneous breaking of chiral symmetry leading to generation of QCD quark mass in addition to the EW mass term~\cite{Muller2005}, i.e. current free quarks are transformed in constituent confined quarks in hadrons.
We can also note that, in a thermalized medium, the hadrons produced are mainly pions ($\pi^\pm$ and $\pi^0$) which are the only hadrons with a mass slightly below the QCD temperature transition, and they decay in a short time scale: $\tau_{\pi^0} \approx 10^{-16}$~s and $\tau_{\pi^\pm} \approx 10^{-8}$~s.

\subsection{Heavy-ion collisions}\label{subsec-HIC}

The study of the QCD phase diagram can be achieved experimentally by conducting collisions of heavy-ions accelerated at relativitic energies. 
Such a world wide program is at work since more than four decades.
After a serie of experiments in a fixed target mode, limiting the energy density which can be reached, the collider mode was explored from 2000 by the Relativistic Heavy-Ion Collider (RHIC) at BNL and from 2010 by the Large Hadron Collider (LHC) at CERN (see Ref.~\cite{SchukraftStock2015} for the history of heavy-ion physics at CERN). 
On one side, the RHIC is able to perform a beam energy scan with the goal to search for the critical point.
The range explored up to now was $\sqrt{s_{\rm NN}} = 7.7-200$~GeV~\footnote{In heavy-ion collisions, when two nuclei collide, the relevant energy is the nucleon-nucleon centre-of-mass energy $\sqrt{s_{\rm NN}}$.} in Au-Au collisions, corresponding to $\mu_{\rm B} = 20-400$~MeV~\cite{RHICLHC2015}.
On the other side, the LHC reaches the frontier energy with $\sqrt{s_{\rm NN}} = 2.76$~TeV during the run 1 and $\sqrt{s_{\rm NN}} = 5.5$~TeV as ultimate goal in Pb-Pb collisions.

\begin{figure}
\begin{minipage}{0.5\linewidth}
\centerline{\includegraphics[width=0.8\linewidth]{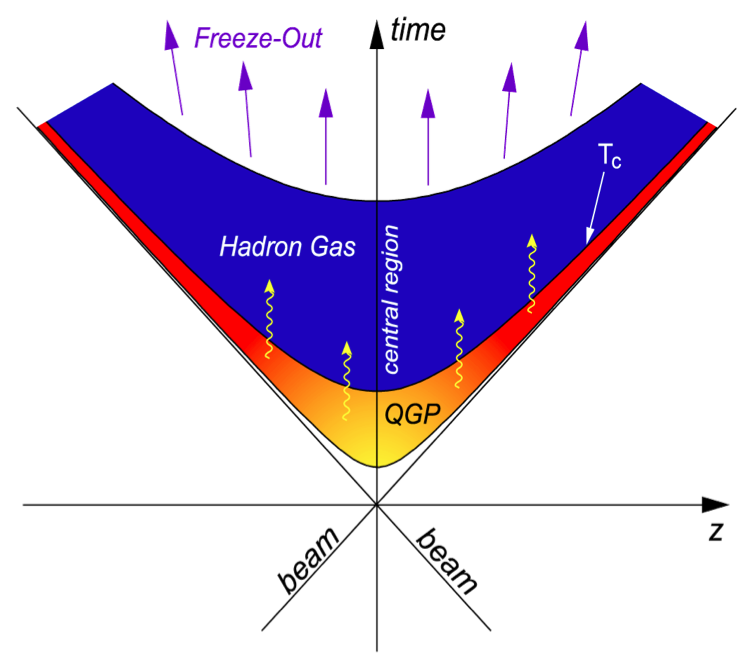}}
\end{minipage}
\hfill
\begin{minipage}{0.5\linewidth}
\centerline{\includegraphics[width=0.8\linewidth]{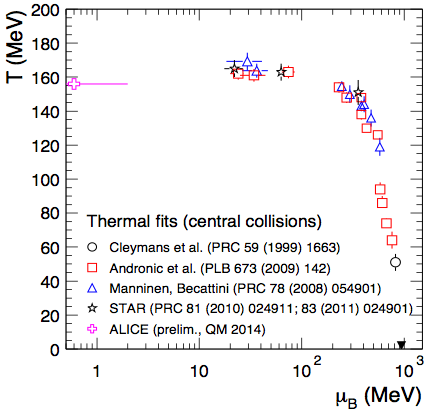}}
\end{minipage}
\caption[]{Left: sketch of the space-time evolution (with only the longitudinal dimension $z$ corresponding to the beam direction) of an ultra-relativistic heavy-ion collision.
Right: compilation in the QCD phase diagram of thermal fit results in central collisions at different energy~\cite{Andronic2014}.
The LHC point is characterized with $\mu_{\rm B}=0$, here reported at $\mu_{\rm B}=0.6$~MeV to be represented in logarithmic scale.}
\label{fig-HIC}
\end{figure}

The time evolution of ultra-relativistic heavy-ion collisions, as those created at RHIC and LHC, can be depicted in four steps, as illustrated in Fig.~\ref{fig-HIC} (left). 
\begin{enumerate}
\item 
In the collision centre-of-mass frame, corresponding to the laboratory frame in Au-Au collisions at RHIC and Pb-Pb collisions at LHC, the two nuclei can be viewed as two pancakes, due to the Lorentz contraction factor, colliding more or less centrally.
\item 
First moment of the collision is characterized by hard (high-energy) processes between incoming quarks and gluons inside the nucleons.
\item 
Next, the bulk thermalizes in a deconfined state, the quark-gluon plasma (QGP), and cooled down in its expansion.
\item 
When the QGP reaches the QCD transition temperature, quarks and gluons hadronize to form an hadronic gaz dominated by pions (step called chemical freeze-out).
Finally, hadrons cease to interact (end of inelastic interactions is called kinetic freeze-out), and they can fly freely to reach the detector elements or decay in meantime.
\end{enumerate}
Produced particles are used to characterize the different steps of the collision.
Below are mentionned the main characteristics of Pb-Pb collisions at LHC; comprehensive reviews can be found in Ref.~\cite{MullerSchukraftWyslouch2012} and~\cite{Andronic2014}.
The measured hadron rates are fitted by a thermal model using three free parameters~\cite{Andronic2014}: the volume of the system at chemical freeze-out $V \approx 5300$~fm$^3$, its temperature $T \approx 156$~MeV and the baryon chemical potential $\mu_{\rm B} \approx 0$.
So, the LHC thermodynamic conditions, reported on the QCD phase diagram of Fig.~\ref{fig-HIC} (right), are very close to the early Universe one.
The Bose-Einstein correlation of identical particles (called also Hanbury-Brown-Twiss or HBT interferometry) is used to measure the size of the system at kinetic freeze-out and its lifetime since the collision: the measurement gives~\cite{ALICEcorrelations2011} $\tau \approx 10$~fm/c $\approx 3 \times 10^{-23}$~s.
Collective behaviours is the sign that the hot and dense matter created in the collision thermalizes very quickly.
These properties can be described by hydrodynamic models and then used to estimate the shear viscosity to entropy density ratio $\eta / s$ of the QGP, measured values are closed to the conjectured AdS/CFT quantum limit~\cite{ShenHeinz2015} $\eta / s \ge 1/4\pi$, and to constrain the EoS state of the QCD matter~\cite{Pratt&al2015}. 
During the QGP phase, hard quarks and gluons produced initially interact with the colored deconfined medium (the QGP) leading to the phenomenon of energy loss (also called jet quenching) due to collisionnal and radiative (gluonstrahlung) processes. 
This effect is quantified by comparing the production of hadrons to the same production in proton-proton collisions where no QGP is expected (this ratio is called nuclear modification factor).
The comparison of hadron energy loss (but also jet modification) to model predictions allows to extract the transport coefficients of quarks and gluons in the QGP.
On the other side, electroweak bosons ($W^\pm$, $Z$ and $\gamma$), which are not sensitive to strong interaction, exhibit no energy loss.  
Furthermore, the energy (more precisely the transverse momentum $p_{\rm T}$) spectrum of direct photons from the bulk is measured and fitted with a black body distribution type to extract the effective temperature of the bulk~\cite{ALICEphotons2015} $T_{\rm eff} \approx 300$~MeV.
In addition, most of the above observations are not seen in p-Pb collisions where the QGP is not expected to be formed, confirming the conclusion that thermodynamic conditions to form a QGP are reached in Pb-Pb collisions.

In summary, present data suggest that the matter created at the RHIC and the LHC behaves like a near-perfect QGP fluid, also referred to as strongly coupled QGP (sQGP).
Furthermore, a first modeling description of all steps involved in heavy-ion collisions is achieved, meaning that heavy-ion physics is on the road towards the development of a Little Bang Standard Model~\cite{Heinz2013}.

\section{The QCD phase transition in early Universe}

The link between the QCD phase transition and the early Universe is done via the Friedman equation involving the Hubble parameter $H(t)$, i.e. the cosmological scale factor $a(t)$
\begin{eqnarray}
H^2 = \left(\frac{\dot{a}}{a}\right)^2 = \frac{8\pi G_{\rm N}}{3} \rho
\label{eq-Friedman}
\end{eqnarray}
with $G_{\rm N}$ the gravitational constant and $\rho$ the mass density of the Universe connected to its energy density by $\epsilon = \rho c^2$.
Assuming that the {\it particle era} of the early Universe is dominated by ultra-relativistic particles~\footnote{(i) Last Planck results~\cite{Planck} show that the Universe is flat, i.e. the curvature term vanishes in the Friedman equation. (ii) The contribution of the dominant cosmological constant in the present Universe was negligible in the early Universe. (iii) At high temperature in the early Universe, the contribution of non-relativistic matter in the energy density is expected to be negligible.}, the energy density is of radiation type $\epsilon_{\rm rad} = \frac{\pi^2}{30} g_T T^4$, where $g_T$ is the number of d.o.f. at temperature $T$. 
With this assumption, we deduce that the cosmological time (in s), related to the Hubble constant by $t = \frac{1}{2H}$ for radiation content, can be written as a function of the Universe temperature (in MeV)
\begin{eqnarray}
t_{\rm [s]} = \frac{2.42}{\sqrt{g_T}(T_{\rm [MeV]})^2}
\label{eq-time}
\end{eqnarray}

\begin{figure}[h]
\begin{minipage}{0.5\linewidth}
\centerline{\includegraphics[width=0.9\linewidth]{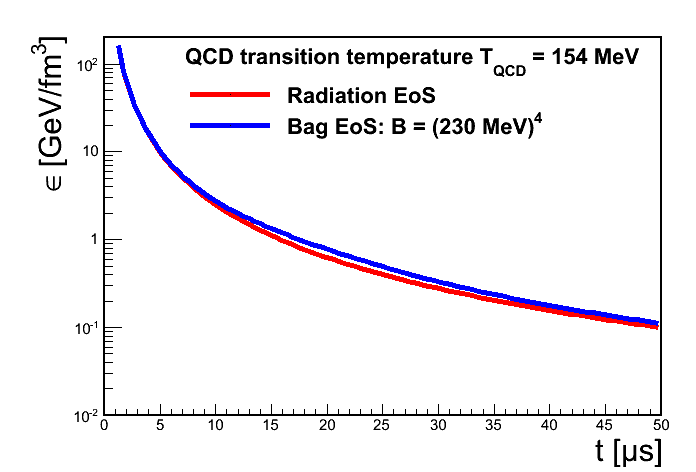}}
\end{minipage}
\hfill
\begin{minipage}{0.5\linewidth}
\centerline{\includegraphics[width=0.9\linewidth]{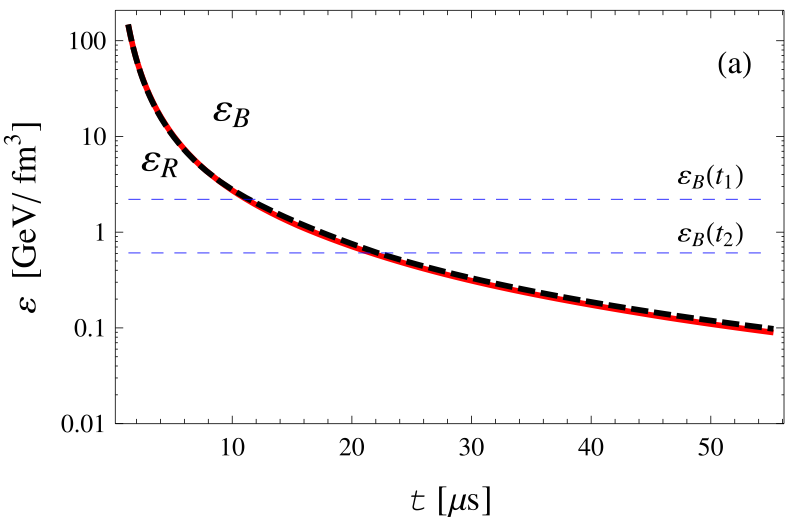}}
\end{minipage}
\begin{minipage}{0.5\linewidth}
\centerline{\includegraphics[width=0.9\linewidth]{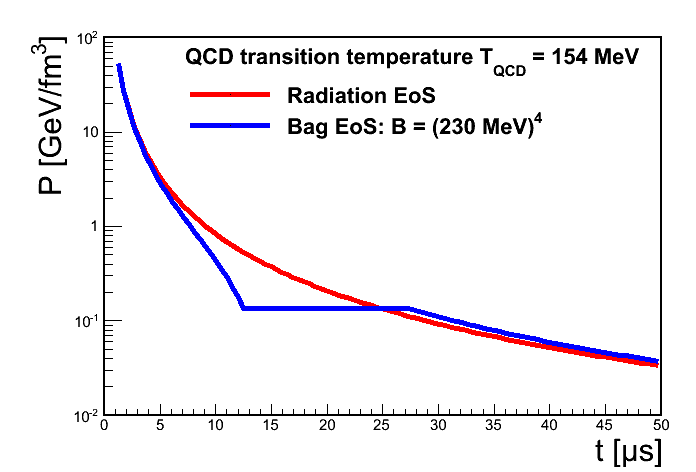}}
\end{minipage}
\hfill
\begin{minipage}{0.5\linewidth}
\centerline{\includegraphics[width=0.9\linewidth]{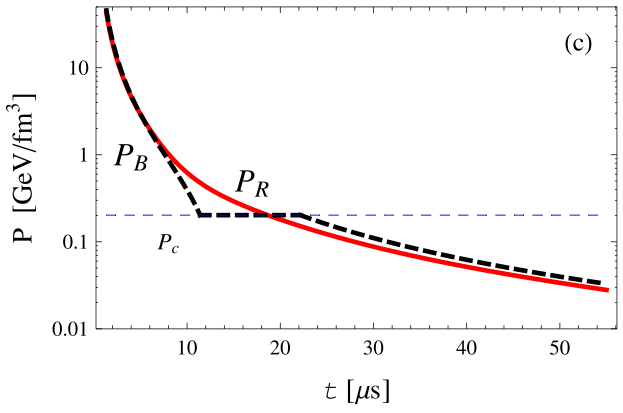}}
\end{minipage}
\begin{minipage}{0.5\linewidth}
\centerline{\includegraphics[width=0.9\linewidth]{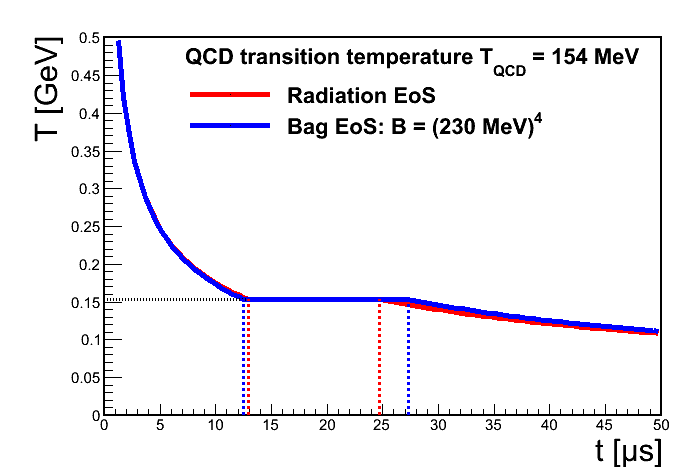}}
\end{minipage}
\hfill
\begin{minipage}{0.5\linewidth}
\centerline{\includegraphics[width=0.9\linewidth]{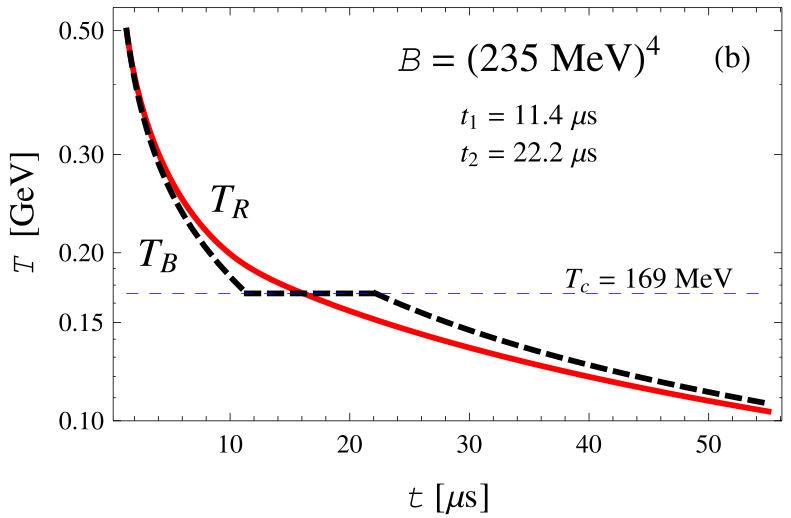}}
\end{minipage}
\caption[]{Time evolution of the energy density $\epsilon$ (upper plots), the pressure $P$ (middle plots) and the temperature $T$ (bottom plots) of the early Universe.
In left plots a 1$^{\rm st}$ order QCD phase transition is assumed $T_{\rm QCD} = 154$~MeV: for a purely radiation EoS (red line) and for the bag EoS (blue line).
Right plots~\cite{Florkowski2011} compare the result of the bag EoS with $T_{\rm QCD} = 169$~MeV (denoted $B$, black dashed line) and a more realistic cross-over transition (denoted $R$, red line).}
\label{fig-EPT}
\end{figure}

A naive approach of the QCD phase transition consists of assuming a first order transition at the temperature predicted by lQCD, $T_{\rm QCD} = 154$~MeV, between two phases of a Universe filled of ultra-relativistic particles, i.e. particles with a mass smaller than the temperature.
So, for $T > T_{\rm QCD}$ (and $T < 1$~GeV), the radiative Universe is characterized by $g_T \approx 62$, taking into account photon, gluons (8), leptons (2 charged and 3 neutrinos) and quarks ($u$, $d$ and $s$) d.o.f.
We can note that most of the d.o.f. (47.5) come from quarks and gluons, we can then qualified this phase of QGP.
For $T < T_{\rm QCD}$ (and $T > m_\pi$), the radiative Universe is characterized by $g_T \approx 17$ with photon, pions ($\pi^\pm$ and $\pi^0$) and leptons d.o.f.
From those numbers of d.o.f. we deduce that the QCD transition starts at $t_i \approx 13\;\mu$s and ends up at $t_f \approx 25\;\mu$s, so with a duration $\Delta t_{\rm QCD} \approx 12\;\mu$s of the order of the cosmological time since the Big Bang.
The time evolution (between 1~$\mu$s and 50~$\mu$s) of the energy density, the pressure and the temperature with this radiation EoS is shown in Fig.~\ref{fig-EPT} (left plots, red line).

To take into account basic QCD properties, as described by the bag model where free massless quarks and gluons are bounded by a negative pressure called the bag constant $B$, the EoS is modified in the following way.
Before the QCD transition, $\epsilon_B = \epsilon_{\rm QGP} + \epsilon_{\rm EW}$ where $\epsilon_{\rm QGP} = \frac{\pi^2}{30} g_{\rm QGP} T^4 + B$ with $g_{\rm QGP} = 47.5$ and $\epsilon_{\rm EW} = \frac{\pi^2}{30} g_{\rm EW} T^4$ with $g_{\rm EW} = 14.25$; while for $T < T_{\rm QCD}$, $\epsilon_B = \epsilon_{\rm HG} + \epsilon_{\rm EW}$ where $\epsilon_{\rm HG} = \frac{\pi^2}{30} g_{\rm HG} T^4$ with $g_{\rm HG} = 3$, assuming that the Hadron Gas (HG) is dominated by pions.
The transition between these two phases is of first order with $\epsilon_B = [1-f(t)] \epsilon_{\rm QGP} + f(t) \epsilon_{\rm HG} + \epsilon_{\rm EW}$ and $0 \le f(t) \le 1$.
The bag constant is constrained by the QCD parameters: $B = \frac{\pi^2}{90} (g_{\rm QGP} - g_{\rm HG}) T_{\rm QCD}^4$.
The time evolution of the energy density, the pressure and the temperature~\cite{YagiHatsudaMiake2005} with this bag EoS is shown in Fig.~\ref{fig-EPT} (left plots, blue line).
The comparison with the radiation EoS shows that the energy density and the temperature have similar shapes, while the pressure exhibits a plateau with the bag EoS. 
Note that the bag constant results in a constant term in the Friedman equation~\ref{eq-Friedman}, like the cosmological constant.

Finally, a more realistic approach was proposed recently to account for the expected cross-over behaviour of the QCD phase transition.
The starting point is the lQCD prediction constrained by the RHIC experimental data~\cite{Florkowski2011}. 
The consequence of this cross-over transition, compared to the previous described $1^{\rm st}$ order phase transitions, is the absence of a constant plateau in the time evolution of pressure and temperature, as shown in Fig.~\ref{fig-EPT} (right plots, red line).

\section{Cosmological implications}

\begin{figure}[h]
\begin{minipage}{0.5\linewidth}
\centerline{\includegraphics[width=0.9\linewidth]{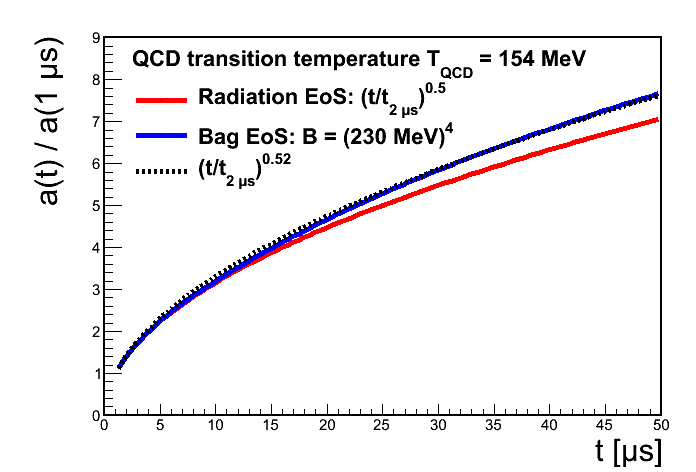}}
\end{minipage}
\hfill
\begin{minipage}{0.5\linewidth}
\centerline{\includegraphics[width=0.9\linewidth]{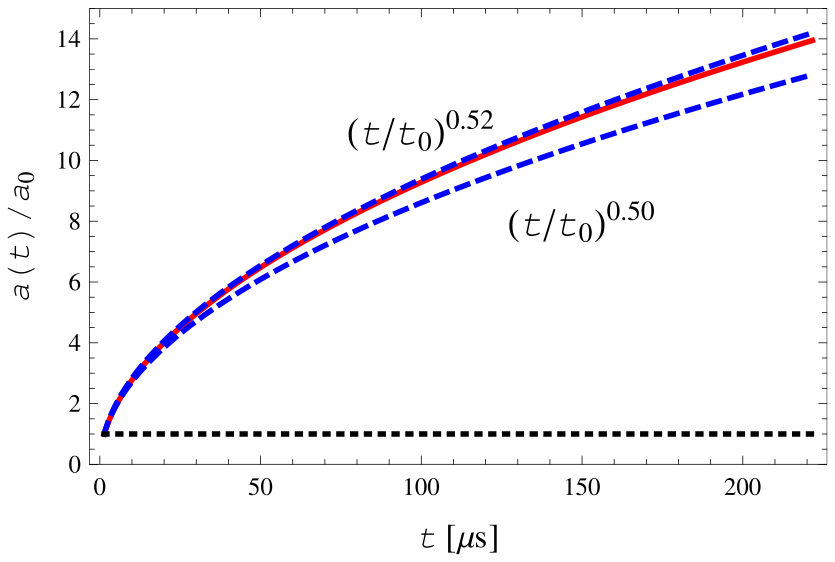}}
\end{minipage}
\caption[]{Time evolution of the cosmological scale factor $a(t)$ with respect to its initial value $a_0 = a(t_0)$ for different QCD phase transition hypothesis.
Left: comparison between the radiation EoS (red line) and the bag EoS (blue line) with $t_0 = 1\;\mu$s and $T_{\rm QCD} = 154$~MeV.
The black dashed line represents a trend of the form $(t/t_0)^{0.52}$.
Right~\cite{Florkowski2011}: the cross-over EoS (red line) is compared to the radiation EoS (lower blue dashed line) and a trend of the form $(t/t_0)^{0.52}$ (upper blue dashed line), with $t_0 = 1.35\;\mu$s and $T_{\rm QCD} = 169$~MeV.}
\label{fig-A}
\end{figure}

Integration of Eq.~\ref{eq-Friedman} leads to the time evolution of the cosmological scale factor 
\begin{eqnarray}
a(t) = a(t_0) \exp\left[\int_{t_0}^t \sqrt{\frac{8\pi G_{\rm N} \epsilon (t^\prime)}{3c^2}}dt^\prime\right] 
\label{eq-scalefactor}
\end{eqnarray}
A purely radiation EoS gives the standard behaviour $a(t)/a(t_0) = \sqrt{t/t_0}$.
In Fig.~\ref{fig-A}, this trend is compared to the bag EoS (left plot) and the cross-over EoS (right plot).
In both cases, even if inputs are not exactely the same (different values of $t_0$ and $T_{\rm QCD}$), the scale factor related to the initial one, i.e. the ratio $a(t)/a(t_0)$, starts to diverge of the radiation EoS behaviour at the level of the QCD phase transition.
And quite surprisingly a similar trend of the form $(t/t_0)^{0.52}$ is observed for both EoS despite the different QCD phase transition assumed: $1^{\rm st}$ order for bag EoS and cross-over for the realistic one.

Beyond this impact on the time dependence of global thermodynamic quantities and the cosmological scale factor, the QCD phase transition can also affect the density fluctuations in the early Universe.
The first study of the energy density fluctuations was carried out for the bag EoS~\cite{SchmidSchwarzWiderin1997} and have shown that the effect of the QCD transition is an amplification of the intial fluctuations.
The source of this amplification is the vanishing value of the sound speed during the $1^{\rm st}$ order phase transition.
The more recent study with a realistic cross-over EoS~\cite{Florkowski2011} shows an opposite trend: a decrease of the QCD energy density fluctuations by about 30\% during the first 220~$\mu$s, and in the same time a triggering of the electroweak energy density fluctuations (assuming no initial fluctuations of this component).

\section{Conclusion}

A review of the QCD phase transition was presented from a theoretical and an experimental points of view with empahasis on its cosmological implications.
Whatever is the nature of this transition, a simple modeling of the early Universe content shows that the time evolution of global thermodynamic quantities does not follow the radiation EoS expectations.
Furthermore, the cosmological scale factor is affected by this {\it particle era} with a time dependence $(t/t_0)^p$ characterized by a power parameter close to $p=0.52$, instead of the classical $p=0.5$ value of the radiation law.
Less clear is the influence of the QCD phase transtion on the initial density fluctuations which can be affected in opposite trend following the nature of the transition.

In summary, it appears that the knowledge of the nature of the QCD phase transition is a fundamental scientific question.
Indeed, beyond the knowledge of the properties of the strong interaction in extreme thermodynamic conditions, the QCD phase transition can play an important role in the understanding of the early Universe. 
This extrapolation must be done keeping in mind that in accelerator experiments: (i) initial state comes from two high energy heavy nuclei, compared to the thermalized cooled down particle bath with initial fluctuations of the early Universe; (ii) the QGP expansion occurs in a pre-exisiting Euclidean 3D space, with respect to the 4D cosmological space-time evolution based on General Relativity; (iii) the time scale involved in accelerator experiments is several orders of magnitude smaller (about $10^{-18}$) than the time scale characterizing the QCD phase transition in the early Universe. 

\section*{Acknowledgments}

The autor thanks the organizers of the {\it Rencontre du Vietnam 2015 on Cosmology} for giving the opportunity to discuss an exotic topic for this conference which involves a convergence of both Physics scales.
The autor would also like to thank Philippe Crochet and Anton Andronic for fruitful discussions on QCD and heavy-ion physics. 


\section*{References}

\begin{thebibliography}{99}

\bibitem{LHC-BEH} ATLAS Collaboration, Phys. Lett. B 716 (2012) 1-29. \\ CMS Collaboration, Phys. Lett. B 716 (2012) 30-61.

\bibitem{Planck} Planck Collaboration, Astron. Astrophys. 571 (2014) A1. \\ Planck Collaboration, arXiv:1502.01582 [astro-ph.CO].

\bibitem{PDG2015} Particle Data Groupe, LBNL  2015, http://pdg.lbl.gov.

\bibitem{NRC2003} {\it Connecting Quarks with the Cosmos: Eleven Science Questions for the New Century}, US National Research Council, http://www.nap.edu/catalog/10079.html.

\bibitem{Ellis2005} J.~Ellis, arXiv:astro-ph/0504501.

\bibitem{Schwarz2003} D.J.~Schwarz, Ann. Phys. (Leipzig) 12, No. 4 (2003) 220-270.

\bibitem{CabibboParisi1975} N.~Cabibbo and G.~Parisi, Phys. Lett. B 59 (1975) 67.

\bibitem{Shuryak1978} J.C.~Collins and M.J.~Perry, Phys. Rev. Lett. 34 (1975) 1353.

\bibitem{CollinsPerry1975} E.~Shuryak, Phys. Lett. B 78 (1978) 150.

\bibitem{MinkowskiKbana2002} P.~Minkowski and S.~Kabana, J. Phys. G: Nucl. Part. Phys. 28 (2002) 2063-2067.

\bibitem{HotQCD2012} Hot QCD Collaboration, Phys. Rev. D 85, 054503 (2012).

\bibitem{HotQCD2014} Hot QCD Collaboration, Phys. Rev. D 90, 094503 (2014).

\bibitem{Muller2005} B.~M\"uller, Nucl. Phys. A 750 (2005) 84.

\bibitem{RHICLHC2015} Y.~Akiba et al., arXiv:1502.02730.

\bibitem{SchukraftStock2015} J.~Schukraft and R.~Stock, arXiv:1505.06853 [nucl-ex].

\bibitem{MullerSchukraftWyslouch2012} B.~M\"uller, J.~Schukraft and B.~Wyslouch, Annu. Rev. Nucl. Part. Sci. 2012.62:361-386.

\bibitem{Andronic2014} A.~Andronic, Int. J. Mod. Phys. A, Vol. 29, No. 22 (2014) 1430047.

\bibitem{ALICEcorrelations2011} ALICE Collaboration, Phys. Lett. B 696 (2011) 328.

\bibitem{ShenHeinz2015} C.~Shen and U.~Heinz, Nuclear Physics News, Vol. 25, No.2 (2015) 6-11.

\bibitem{Pratt&al2015} S.~Pratt et al., Phys. Rev. Lett. 114 (2015) 202301.

\bibitem{ALICEphotons2015} ALICE Collaboration, arXiv:1509.07324 [nucl-ex].

\bibitem{Heinz2013} U.~Heinz, J. Phys. Conf. Ser. 455 (2013) 012044, arXiv:1304.3634 [nucl-th].

\bibitem{YagiHatsudaMiake2005} K.~Yagi, T.~Hatsuda and Y.~Miake, Camb. Monogr. Part. Phys. Nucl. Phys. 23 (2005).

\bibitem{Florkowski2011} W.~Florkowski, Nucl. Phys. A 853 (2011) 173-188.

\bibitem{SchmidSchwarzWiderin1997} C.~Schmid, D.J.~Schwarz and P.~Widerin, Phys. Rev. Lett. 78 (1997) 791-794.

\end{thebibliography}
\end{document}